\documentclass[twocolumn,preprintnumbers,nofootinbib]{revtex4}
\usepackage{graphicx}
\usepackage{amssymb}
\usepackage{color}
\usepackage{enumerate}
\def\beq{\begin{equation}}
\def\eeq{\end{equation}}
\def\beqn{\begin{eqnarray}}
\def\eeqn{\end{eqnarray}}

\renewcommand{\texttt}{{}}
\newcommand{\be}{\begin{eqnarray}}
\newcommand{\ee}{\end{eqnarray}}

\begin{document}

\title{Towards a finite quantum supergravity}
%

\author{Leonardo Modesto}
\affiliation{Perimeter Institute for Theoretical Physics, 
31 Caroline St., Waterloo, ON N2L 2Y5, Canada}

\date{\small\today}

\begin{abstract} \noindent
In this paper we study an $N=1$ supersymmetric extension of 
a perturbatively super-renormalizable (nonlocal)theory of gravity in four dimensions.
The nonlocal supergravity theory is power-counting super-renormalizable and 
tree level unitary with the same particle content of the local $N=1$ supergravity
(as simple example, unitarity of the three dimensional $N=1$ and $N=2$ supergravity is proved). 
We believe that extended $SO(N)$ supergravity, for $N=4$ or $N=8$, might be free from divergences 
also at one loop. The extended supergravities  would then result finite at any order in the loop expansion.

\end{abstract}

\maketitle


One of the greatest revolutions in quantum field theory is the
discovery of supersymmetry. 
In this letter we consider a nonlocal extension of the higher-derivative 
supergravity, and we prove that the theory only includes the 
particles compatible with the individualized graded supersymmetry.
The approach we follow here is the same as that introduced 
in some recent papers, with the limits of $D=4$ 
\cite{BM, modesto, modestoshort, modestoNC, ModestoMoffatNico, Moffat, MF2, MF3, MF4, MF5, MF6, Bis2, Bis3, NS1, NS2, NS3, NS4} 
and a multidimensional spacetime in \cite{ModeMulti}.
The theory fulfills 
a synthesis of minimal requirements: 
(i) classical solutions must be singularity-free;  
(ii) local Supergravity should be a good approximation of the theory at an energy scale $E << M_P$, where 
$M_P$ is the Planck mass;  
(iii) the spacetime spectral dimension has to decrease with the energy
\cite{Spectral}; 
(iv) the quantum theory has to be perturbatively finite or renormalizable;
(v) the theory has to be unitary, with no other 
pole in the propagator in addition to the supergravity multiplet. 

Let us start with the simple $N=1$ supergravity in four spacetime dimensions.
The supersymmetric  multiplet consists of the spin-$2$ graviton, the spin-$3/2$ gravitino and three auxiliary
fields. 
The theory we are going to propose in this paper is a nonlocal extension of the quadratic supergravity 
suggested in  \cite{VNS, FerraraExt},
and it 
has the following structure, 
 \be
&& \hspace{-1cm}
 \mathcal{L} 
 = - \frac{\gamma}{2}\underbrace{(\kappa^{-2}\, R +  {\rm more})}_{\mathcal L_{R} }+ 
  \underbrace{( \kappa^{-2} R \, \alpha(\Box)\,R + {\rm more})}_{{\mathcal L}_{R^2 }} \nonumber \\
&& \hspace{-0.3cm} +  \underbrace{ \kappa^{-2} \left(R_{\mu \nu} \, \beta(\Box) R^{\mu\nu} - \frac{1}{3} R \, \beta(\Box)\,R \right) + {\rm more}}_{{\mathcal L}_{C^2} },
\label{KinS}
 \ee
 where $\kappa^2 = 8 \pi G$.
The word ``more" indicates terms with gravitinos and auxiliary fields necessary to make
each term of the theory locally supersymmetric. On the other hand,  the two ``form factors" 
$\alpha(\Box)$ and $\beta(\Box)$
are ``entire functions" of the covariant D'Alembertian operator 
\cite{efimov, Krasnikov, Tombo}.
For $N=1$ the Super-Poincar\'e Lagrangian density reads 
\be
&& \mathcal{L}_R = \kappa^{-2} R + \bar{\psi} \cdot R + \frac{2}{3} \left(S^2 +P^2 - A_{\mu}^2   \right) \, ,
\nonumber \\
&&   \bar{\psi} \cdot R = e^{-1} \bar{\psi}_{\mu} \, \epsilon^{\mu\nu\rho \sigma} \, \gamma_5 \gamma_\nu 
\mathcal{D}_{\rho} \psi_{\sigma}.
\label{LR}
\ee
As far as the kinetic terms are concerned, the $R^2$-type invariants are: 
\be
&& \mathcal{L}_{R^2} = \kappa^{-2} R \alpha(\Box) R + 
\bar{R} \cdot \gamma \not\!\partial \, \gamma \, \alpha(\Box) \cdot R \nonumber \\
&&- 4 \, (\partial_{\mu}  S) \alpha(\Box) (\partial^{\mu}  S)
- 4 \, (\partial_{\mu}  P) \alpha(\Box) (\partial^{\mu}  P) \nonumber \\
&&+ 4 \, (\partial_{\mu}  A^{\mu}) \alpha(\Box) (\partial^{\nu}  A^{\nu}) ,
\label{LR2}
\ee
and the linearized nonlocal generalization of the super-conformal terms are: 
\be
&& \mathcal{L}_{C^2} = 
\kappa^{-2} \left(R_{\mu \nu} \, \beta(\Box) R^{\mu\nu} - \frac{1}{3} R \, \beta(\Box)\,R \right) \nonumber \\
&& - \bar{\psi}_{\mu} (\Box \delta_{\mu \nu} - \partial_\mu \partial_\nu) \beta(\Box) 
\left(R^{\nu} - \frac{1}{3} \gamma^\nu \gamma \cdot R \right) \nonumber \\
&& - \frac{1}{3} F_{\mu \nu} \beta(\Box) F^{\mu \nu}, 
\label{LC2}
\ee
where $S$ is a scalar term and $P$ is a pseudo-scalar, $A_{\mu}$ is a vector and $F_{\mu \nu}$ its curvature.
Studying unitarity, we will show later that the ``linearized" nonlocal theory respects global supersymmetry, since 
all the states fill up the $N=1$ supergravity multiplet. 
Of course, 
other terms with gravitinos, auxiliary fields and vector fields 
as well as $4$-fermions interactions might be added to implement local supersymmetry \cite{CompleteN1}. 
In this paper we assume the $R^2$ action to be fully known, its nonlocal extension included.
However, these terms are not essential to get the propagators, 
nor to extract information about the spectrum as well as the unitarity of the theory.
The mass dimension of the couplings and fields are: $[\gamma]= 0$, $[\kappa]=-2$, 
$\alpha(\Box)] = [\beta(\Box)] = - 2$, $[\psi_\mu] = 3/2$, $[S] = [P] = [A_{\mu}] = 2$, 
$[R^{\mu} ] = 5/2$.

To calculate the two-point functions,
we first expand the Lagrangian (\ref{KinS}) in powers of the graviton field defined 
by $g_{\mu \nu} = \delta_{\mu \nu} + \kappa \, h_{\mu\nu}$. Then, inverting each quadratic term of the gauge fixed action \cite{VNS}, we get: 
\be
&& \langle h \,h \rangle = \frac{P^2}{\frac{\gamma \Box}{8} + \frac{\beta(\Box) \Box^2}{4} } -
\frac{P^{0,s}}{ \frac{\gamma \Box}{4} - 3 \alpha(\Box) \, \Box^2 } \, , \nonumber \\
&& \langle \psi \, \psi \rangle = - \frac{P^{3/2}}{\not\!\partial \left( \beta(\Box) \, \Box + \frac{\gamma}{2} \right)} + 
\frac{P^{0,s}}{ \not\!\partial \left(   -12 \, \alpha(\Box) \Box + \gamma \right)  } \, , \nonumber \\
&& \langle A \, A \rangle = \frac{P^1}{\frac{\gamma }{3} + \frac{2 \beta(\Box) \Box}{3} } +
\frac{P^{0}}{ \frac{\gamma}{3} - 4 \alpha(\Box) \, \Box } \, , \nonumber \\
&& \langle S \, S \rangle = \langle P \, P \rangle = \frac{1}{4 \alpha(\Box) \Box - \frac{\gamma}{3}} \, .
\label{props}
\ee
In (\ref{props}) all the indexes have been omitted and  
the projectors that satisfy orthonormality, decomposition of the unity and completeness \cite{VN}, are: 
\be
&& P^{3/2}_{\mu \nu} = \theta_{\mu \nu} -  \frac{1}{3} \hat{\gamma}_{\mu} \hat{\gamma}_{\nu} \,  ,
\hspace{0.2cm}
\hat{\gamma}_{\mu} = {\gamma}_{\mu} - \omega_{\mu} \, , 
 \nonumber \\
&& (P^{1/2}_{11} )_{\mu \nu} = \frac{1}{3} \hat{\gamma}_{\mu} \hat{\gamma}_{\nu} \,  ,
\hspace{0.2cm} 
 (P^{1/2}_{12} )_{\mu \nu} = \frac{1}{\sqrt{3}} \hat{\gamma}_{\mu}\omega_{\nu} \,  ,
 \nonumber \\
 &&
 (P^{1/2}_{21} )_{\mu \nu} = \frac{1}{\sqrt{3}} \omega_{\mu} \hat{\gamma}_{\nu} \,  ,
 \hspace{0.2cm}
 (P^{1/2}_{22} )_{\mu \nu} = \omega_{\mu}\omega_{\nu} \,  , \nonumber \\
 &&
  \theta_{\mu \nu} = \delta_{\mu \nu} - \omega_{\mu} \omega_{\nu} \,  ,
  \hspace{0.2cm} 
  \omega_{\mu} = \frac{\partial_{\mu} \not\!\partial}{\Box} \, .
\ee
$P^{2}, P^{0,s}, P^{0,ts}$ 
are the spin-$2$ projectors defined in \cite{VN} and $P^1$ and $P^0$ the vector field projectors.
Let us introduce the following notations:
\be
\alpha(\Box) := \alpha_0 + h_0(\Box) \, , \hspace{0.2cm} 
\beta(\Box) := \beta_2 + h_2(\Box) \, , 
\label{alphabeta}
\ee
where $h_2(z)$ and $h_0(z)$ are two entire functions of their argument $z = \Box$,
which can be defined in terms of a
single entire function $H(z)$,
\be
&& h_2(z) = \frac{ \tilde{\gamma} (e^{H(z)} -1 ) -  2\, \tilde{\beta}_2 \, z }{2 \, z}, \nonumber \\
&& h_0(z) = - \frac{ \tilde{\gamma} (e^{H(z)} -1 ) +  12 \, \tilde{\alpha}_0 \, z }{12 \, z}. 
\label{h2h0}
\ee
%
%
We can express the graviton propagator as follows, 
\be
&& \langle h \, h \rangle = \frac{8 P^2}{ \Box \bar{h}_2 (\Box) }  
- \frac{4 P^{0,s}}{ \Box \bar{h}_0 ( \Box) } \, , \label{Vz}\\
&& \bar{h}_2(\Box) \equiv \frac{\gamma}{8} + \frac{\beta(\Box) \Box}{4} \, , \nonumber \\ 
&& \bar{h}_0(\Box) \equiv  \frac{\gamma }{4} - 3 \alpha(\Box) \, \Box \, .  \nonumber 
\ee
Given the above $2$-point function, we require 
$\bar{h}_2(z), \bar{h}_0(z)$ to be real and positive on the real axis and without zeroes on the 
whole complex plane for $|z| < + \infty$. This requirement implies that there are no 
gauge-invariant poles other than the transverse massless physical graviton pole.
A similar requirement has to be satisfied by the other fields in the theory. 

We are going to show that, for a given choice of $\alpha(\Box)$ and $\beta(\Box)$, 
we are able to remove at the same time the extra poles in the graviton, as well as the gravitino and
the auxiliary fields propagators. This will be a positive proof of global supersymmetry, 
since all the states 
fill up the $N=1$ supergravity multiplet \cite{FerraraExt}. 

Let us assume now that the quantum theory is renormalized at some scale $\mu_0$. 
If we want the bare propagators to possess no other gauge-invariant poles 
in addition to those that fill up the $N=1$ supergravity multiplet, and if we want 
in particular 
to avoid ghosts to preserve unitarity, then we have to set  
\be
 {\alpha}_0(\mu_0) = \tilde{\alpha}_0 \,,  \,\,\,\, 
 {\beta}_2(\mu_0) = \tilde{\beta}_2 \, , \,\,\,\, 
 {\gamma}(\mu_0) = \tilde{\gamma} \,  .
\label{betaalpha}
\ee
If we choose another renormalization scale $\mu \neq \mu_0$, then the bare propagators acquire poles; 
however, these poles 
cancel out in the dressed physical propagator with a corresponding shift in the self energy.
Using (\ref{betaalpha}) all the propagators simplify to 
\be
&& \langle h \,h \rangle = 8 \frac{e^{-H(\Box)}}{\gamma \, \Box}  \left[ P^2 - \frac{P^{0,s}}{2} \right]  \, , 
\label{graviton} \\
&& \langle \psi \, \psi \rangle = - 2 \frac{e^{-H(\Box)}}{\gamma \not\!\partial}   
\left[ P^{3/2} - 2 P^{0,s} \right] \, , \label{fermion} \\
&& 
\langle A \, A \rangle = 3 e^{-H(\Box)} 
\left[ P^1 + P^{0} \right] \gamma^{-1} \, ,  \label{Af} \\
&& 
\langle S \, S \rangle = \langle P \, P \rangle = - 3 e^{-H(\Box)} \, \gamma^{-1} 
\label{SPf}\,.
\ee
Since $H(z)$ is an entire function, the pole structure of the $2$-point correlators respects
the $N=1$ global supersymmetry 
and no other 
particle appears in the theory. 
In particular it is only at high energy that the auxiliary fields 
acquire a sort of kinematic term. 
Supersymmetry allows us to maintain unchanged the spectrum with a single choice of 
the operators $\alpha(\Box)$ and $\beta(\Box)$. Once again, this is also a consistency 
check of the truncated Lagrangian 
\be \mathcal{L} \approx\mathcal{L}_R + \mathcal{L}_{R^2}+ \mathcal{L}_{C^2} \,  .
\ee 



Following Efimov's study on nonlocal interactions \cite{efimov},
we identify a class of theories 
characterized by 
the form factor $V(z) : = \exp - H(z)$ in (\ref{h2h0}). 
Let us consider the gauge invariant graviton 
propagator (\ref{graviton}) in the following general form 
(please note that the other propagators (\ref{fermion}-\ref{SPf}) 
are uniquely set once the graviton propagator has been chosen),
\be
D(z) = \frac{V(z)}{ z }
\label{propgeneral}
\ee
(the notation is rather compatible with the graviton propagator 
and $z := \Box$).

As shown by Efimov \cite{efimov}
, the nonlocal field theory is ``unitary" and ``microcausal"
provided that the following properties are satisfied by $V(z)$, 
\begin{enumerate}
\renewcommand{\theenumi}{\Roman{enumi}}
\item
$V(z)$ is an entire analytic function in the complex
$z$-plane and has a finite order of growth $1/2 \leqslant \rho < + \infty$ i.e. $\exists \, b>0,c>0$ so that 
\be
|V(z)| \leqslant c \, e^{b \, |z|^{\rho}}.
\ee
\item When ${\rm Re}(z) \rightarrow + \infty$ ($k^2 \rightarrow + \infty$), $V(z)$ 
decreases with sufficient rapidity. We can encounter the following cases:
\begin{enumerate}
\renewcommand{\labelenumii}{\alph{enumii}.}
    \item $V(z) = O\left(\frac{1}{|z|^a}  \right)$ ($a>1$) , \\
    \item $\lim_{ {\rm Re}(z) \rightarrow +\infty} |z|^N |V(z)| =0$, $\forall \, N>0$. 
\end{enumerate}
\item  $[V(z)]^{*}= V(z^*)$ 
and $V (0) = 1$. The function $V(z)$ can be non-negative on 
the real axis, i.e. $V(x) \geqslant 0$, $x = {\rm Re}(z)$.
\end{enumerate}
We also define the following conical region, 
\be
&& \hspace{-0.2cm} 
C = \{ z \, | \,\, - \Theta < {\rm arg} z < + \Theta \, , \,\,  \pi - \Theta < {\rm arg} z < \pi + \Theta \} , \nonumber \\
&&  \hspace{-0.2cm} 
{\rm for } \,\,\, 0< \Theta < \pi/2. \label{cono} 
\ee
%
Here 
we study the 
II.a. example. 
Once we express the form factor as the exponential of an entire function $H(z)$, 
\be
V(z) = e^{- H(z)} , 
\ee
an example of entire function $H(z)$, which is compatible with the property II.a, is
\be
&& \hspace{-1cm} 
H(z) =  \frac{1}{2} \left[ \gamma_E + 
\Gamma \left(0, p_{a}^{2}(z) \right) \right] + \log [ p_{a}(z) ] \,  ,
\label{H} \\
&& \hspace{-0.1cm}  \equiv \sum_{n =1}^{+ \infty} ( -1 )^{n-1} \, \frac{p_{a}(z)^{2 n}}{2n \, n!} \,\, ,
\hspace{0.2cm} {\rm Re}( p_{a}^{2}(z) ) > 0  \, , 
\nonumber 
\ee
where $p_a(z)$ is a real polynomial of degree $a$, 
$\gamma_E$ 
is the Euler's constant and  
$\Gamma(a,z)$ 
is the incomplete gamma function.  
If we choose $p_{a}(z) = z^{a}$, 
 the $\Theta$ angle, which defines the cone $C$,  
is $\Theta = \pi/4 a$. 
%

Given the above properties, let us study the ultraviolet behavior of the quantum theory.
Following \cite{VN, FerraraExt, CompleteN1} we assume that there are no other 
ultraviolet relevant vertexes to evaluate the power-counting renormalizability
besides the following, 
\be
&& V_{h} \sim h^{\ell}  (\partial^2 h) \, \frac{e^{H(\Box)}}{\Box} \,(\partial^2 h) \,  , \nonumber \\ \,\,\,\, 
&& V_{\psi} \sim h^{\ell} ( \partial \psi) \, \frac{e^{H(\Box)}}{\Box} (\partial^2 \psi) \, ,  \nonumber \\
&& V_S \sim h^{\ell} \, e^{H(\Box)} S^2 \, \, , \,\,\,\, V_P \sim h^{\ell} \, e^{H(\Box)} P^2 \, , \nonumber \\
&& V_A \sim h^{\ell} \, \partial A \, \frac{e^{H(\Box)}}{\Box} \,  \partial A \,\, , \,\,\,\, \nonumber \\
&& V_{4 \psi} \sim h^{\ell}  ( \psi \partial \psi) \, \frac{e^{H(\Box)}}{\Box} \,  ( \psi \partial \psi) \, ,
\label{vertex}
\ee
where $h^{\ell}$ indicates a vertex with $\ell-$gravitons. 
We introduce the following notation: $B_h$, $B_S$, $B_P$, $B_A$, $F_{\psi}$ indicate respectively 
the number of internal lines for the graviton, scalar, pseudo-scalar, vector and fermion fields; 
$V_h, V_S, V_P, V_A, V_{\psi}, V_{4\psi}$ 
indicate the number of vertexes with the structure defined in (\ref{vertex}).
The upper bound to the $L-$loops amplitudes reads,
{\small \begin{eqnarray}
 \hspace{-0.1cm} 
\mathcal{A}^{(L)} \! \leqslant \! 
\int (dk)^{4 L} \left( e^{-H} \right)^{L-1} k^{- \overbrace{[ 2 (B_h - V_h)+ F_{\psi} - V_{\psi}]}^{\geqslant 0} }   \, ,
\label{divera}
\end{eqnarray}}
\hspace{-0.3cm}
In (\ref{divera}) we used again the topological relation between vertexes, internal lines and 
number of loops, which for this theory reads %
\be
&& \hspace{-0.96cm} L - 1 = B_h + B_S + B_P + B_A + F_{\psi} \nonumber \\
&&-  V_h - V_S - V_P - V_A - V_{\psi} - V_{4\psi}  .                
\ee     
For the theory defined by the entire function (\ref{H}) and the minimal choice $p_a(z) = z^2$,
the superficial degree of divergence resulting from the amplitudes (\ref{divera}) is 
\be \,\,\, \,\, 
\delta \leqslant 4 - 2(a-2) (L-1) - [2 (B_h - V_h) + (F_{\psi} - V_{\psi})]. \nonumber 
\ee
%
Therefore, 
only 
1-loop divergences survive if $a>4$ and the theory is super-renormalizable
\cite{shapiro}. 
In this theory,  
 the only quantities still to be renormalized are 
$\gamma$, $\beta_2$, $\alpha_0$. The form factors $\alpha(\Box)$ and $\beta(\Box)$
can be measured experimentally like the analog quantities in experimental particle physics.


%
Here 
we study the 
II.b. example of form factor, 
\be
V(z) = e^{- z^n} \,\, {\rm for} \,\,  n \in \mathbb{N}_+ \, , \,\, 
 \,\,  \rho = n < +\infty. 
\ee
When omitting the tensorial structure, the high energy propagator in the momentum space reads 
 $D(k) = \exp{- (k^2/\Lambda^2)^n}/k^{2}$.
From (\ref{divera}) we see that
the $L$-loops amplitude is UV finite for $L>1$ and it still diverges like ``$k^4$" for $L=1$.


Let us see if we can make the theory convergent also at one-loop.
In other words, we are going to investigate if and under what conditions the theory can be ``finite".
The well-known argument about one-loop finiteness of $N=1$ supergravity develops like in \cite{1loop}.
The one loop counterterms for any matter system ($\Phi$) that is covariantly coupled to Einstein 
gravity are 
{\small \be 
\!\!\! \Delta \mathcal{L} \! = \! \frac{1}{\epsilon} 
\big[x R_{\mu \nu}^2  + y R^2 
+ z  R_{\mu \nu \rho \sigma} M^{\mu \nu \rho \sigma}(\Phi) + w  N(\Phi)  \big]  ,
\label{Div1}
\ee}
\hspace{-0.27cm}
where $\epsilon =  D-4$, 
while $M^{\mu \nu \rho \sigma}(\Phi)$ and $N(\Phi)$ are polynomial in the background
fields $\Phi$, in their derivatives $\partial \Phi$ and in the vierbain $e_{a \mu}$.
In Einstein gravity the equations of motion $R_{\mu\nu} - R g_{\mu\nu}/2 = - \kappa^2 T_{\mu \nu}$
are satisfied and we can express (\ref{Div1}) in the following equivalent way,
{\small  \be 
\Delta \mathcal{L} 
= \frac{1}{\epsilon} 
\left[4 x \kappa^4 T_{\mu \nu}^2  + 4 y \kappa^4 T^2 
+ z R \cdot M(\Phi) 
+ w N(\Phi)  \right] ,
\label{Div2}
\ee}
\hspace{-0.28cm} 
where we defined  
$R \cdot M(\Phi) := R_{\mu \nu \rho \sigma} M^{\mu \nu \rho \sigma}(\Phi)$ and $T$ 
is the trace of the stress-energy tensor.
In this form $\Delta \mathcal{L}$ does not contain terms with only the gravitational field, and 
clearly $\Delta \mathcal{L}$ has vanishing matrix elements between purely graviton states.
On the other hand, global supersymmetry relates the amplitudes with the external matter fields 
to the amplitudes with only external gravitons, since the matter fields are in the 
same multiplet as the supergravitons. Therefore, since $\langle \Delta \mathcal{L} \rangle \equiv0$ on any external state, we infer the theory is one-loop finite.
In our theory the same argument does not apply because the classical Einstein equations 
are no longer satisfied.
Let us explain more in detail this point. 
For $\tilde{\gamma} = \gamma$, $\tilde{\alpha}_0 = \alpha_0$ and $\tilde{\alpha}_2 = \alpha_2$,
the super-renormalizable action simplifies to 
\be
\hspace{-0.2cm} 
\mathcal{L} = 
\frac{\gamma}{2 \kappa^2}  \left[ - R + G^{\mu \nu} 
\left(\frac{e^{H(\Box)} -1}{\Box} \right)R_{\mu\nu} \right] + \mathcal{L}_M \, ,
\label{compact}
\ee
where $G_{\mu \nu}$ is the Einstein tensor and $\mathcal{L}_M$ is the Lagrangian for
the matter fields.
The modified Einstein equations deriving from (\ref{compact}) are
\be
&& E_{\mu \nu} = - \kappa^2 e^{- H(\Box)} T_{\mu \nu} \, , \label{eqm} \\
&& 
E_{\mu \nu} \equiv R_{\mu \nu} - \frac{1}{2} g_{\mu \nu} R + O(R^2) + \dots + O(\Box^{a-1} R^2) \, . 
\nonumber 
\ee
We observe that there are no other linear terms in the curvature besides the Einstein tensor in
(\ref{eqm}). 
If we invert (\ref{eqm}), for the Ricci tensor we find
\be
&& \hspace{-0.5cm}
 R_{\mu \nu} = E_{\mu \nu} - \frac{1}{2} g_{\mu \nu} g^{\alpha \beta} E_{\alpha \beta} + O(R^2)
- \kappa^2 e^{- H(\Box)} T_{\mu \nu} \, , \nonumber \\
&& \hspace{-0.5cm}
R = - g^{\mu \nu} E_{\mu \nu} + O(R^2) - \kappa^2 e^{ - H(\Box)} T_\mu^\mu \, . 
\label{R}
\ee
If we replace $R_{\mu\nu}$ and $R$ inside $\Delta \mathcal{L}$,  
we get other operators $O(R^2)$ which do not
vanish when evaluated between external graviton states. 
On the other hand, the third and forth terms in (\ref{Div2}) 
still have zero expectation value. We conclude that the theory is not
one-loop finite. 

In what follows, we are going to prove 
 the ``uniqueness" of the propagator (\ref{graviton}) 
for the case of a flat fixed background.  
Given the action (\ref{compact}) and the equations of motion (\ref{eqm}), 
we want to prove 
the absence of ghosts and therefore the unitarity 
of the theory. 
We know that at one loop the more general counter terms we can add to the Lagrangian
(\ref{compact}) are those given in (\ref{Div1}), so the regularized Lagrangian takes the 
following form  
\be
\mathcal{L}_{\rm Reg} = \mathcal{L} + \Delta \mathcal{L}.
\label{Lreg}
\ee
Since we are interested in possible modification to the graviton propagator, we consider only 
the first two purely gravitational terms in (\ref{Div1}). 
Using the field equations (\ref{eqm}, \ref{R}) we can rephrase such 
counter terms in the following form,
 \be
  && \hspace{-0.5cm} \Delta {\mathcal L} = \frac{1}{\epsilon} \left(x \, R_{\mu \nu} R^{\mu \nu} + y \, R^2 \right) 
  \nonumber \\&&  \hspace{0.15cm} 
= \frac{1}{\epsilon} \left(   x  \, E_{\mu \nu}  R^{\mu \nu} + \tilde{y} \, g^{\alpha \beta} E_{\alpha \beta}  R + O(R^3) \right), 
\label{Dg}
 \ee
where $\tilde{y} = -(y +1/2)$.
Considering the following covariant redefinition of the metric field 
\be 
\delta g_{\mu \nu} = \frac{x}{\epsilon} \, R_{\mu \nu} + \frac{\tilde{y}}{\epsilon} \,  R \, g_{\mu \nu} 
\label{Rg}
\ee
the divergence (\ref{Dg}) becomes 
\be
\Delta \mathcal{L} = E_{\mu \nu} \, \delta g^{\mu \nu} + O(R^3),
\ee
which is $O(R^3)$ on-shell ($E_{\mu \nu} = 0$).
In the new field variable the Lagrangian reads 
\be
\mathcal{L}+ \Delta \mathcal{L} \rightarrow \mathcal{L} + O(R^3) \, , 
\label{Lreg2}
\ee
showing that the poles of the propagator are left unchanged under quantization.
Since the amplitudes are invariant under a covariant field redefinition, then here we showed 
the uniqueness of the propagator without affecting the renormalizability properties
of the theory. The most relevant point in this proof is the absence of other linear terms in the curvature,  
such as $\nabla \nabla R_{\dots}$ in the classical equations of motion (\ref{eqm}).
Our result is in perfect agreement with the 
gravitational renormalizable theory presented in \cite{ansel}.
In such paper the author elegantly shows that a theory without quadratic terms in the curvature, 
but with infinite number of operators $O(R^3)$, 
is ghost free and renormalizable.

In this paper we showed that $N=1$ nonlocal supergravity has the same spectrum of the local theory.
We believe that the same result also holds for the extended $SO(N)$ supergravity, at least for 
$N\leqslant4$ on the basis of the results in \cite{FerraraExt}.
If this is the case, we will have a theory off-shell finite even at one loop, 
as a consequence of
the Nielsen-Hughes formula \cite{HamberBook}. 
Such formula tells that the one loop beta function of a spin-$s$ particle reads 
\be
\beta_s = - (-1)^{2 s} \, \left[ (2 s)^2 - \frac{1}{3} \right]. 
\label{N-H}
\ee
Because of the global supersymmetry, we can only have quadratic counterterms in the curvature at one loop.
However, for $N=4$ the particle multiplet consists of: a vierbain field 
$e_{a \mu}$, four spin $3/2$ Majorana fields $\psi^i_{\mu}$, six vector fields $A_{\mu}^{ij}$, four $1/2$ Majorana fields 
$\xi^i$, one scalar $A$ and one pseudo scalar $B$. Consequently  
{\small \be 
\hspace{0.2cm} 
\beta = \hspace{-0.7cm} \sum_{s=2,3/2,1,1/2,0} \hspace{-0.6cm} \beta_{s} = - \frac{47}{3} \cdot 1 + \frac{26}{3} \cdot 4 
- \frac{11}{3} \cdot 6 + \frac{2}{3} \cdot 4 +  \frac{1}{3} \cdot 1 = 0 \, . \nonumber 
\ee}
\hspace{-0.27cm}
This result proves that the off-shell $N=4$ theory might be finite. 

In view of the results presented in this paper, we do not a priori exclude the
possibility of defining at least at linear level an higher-derivative $N=8$ supergravity that
includes nonlocal form factors. Indeed, if the entire functions avoid extra poles in the 
propagators, then the argument in \cite{FerraraExt} does not apply and an higher derivative extension 
of $N=8$ supergravity seems constructible. Should that be the case, 
$\beta =0$ and the $N=8$ theory would result to be finite.



\section*{APPENDIX: \\ $N=1$ AND $N=2$ THREE DIMENSIONAL \\ (CURVATURE)$^2$ ``NONLOCAL" SUPERGRAVITY} 
In this section we study a nonlocal extension of the three-dimensional supergravity
\cite{3dSugra}. The particle multiplet of $N=1$ supergravity consists of 
$(e_{\mu}^m , \psi_\mu, S)$, where $e_{\mu}^m$ is the dreibein, $\psi_{\mu}$ is the 
Majorana-gravitino and $S$ a real scalar auxiliary field.
The Lagrangian we propose here consists of the following three terms, 
\be
&& \mathcal{L}_0 = - \frac{e R}{4 \kappa^2} + \frac{1}{2 \kappa^2} 
\epsilon^{\mu \nu \rho} (\bar{\psi}_\mu D_{\nu} \psi_{\rho} ) - \frac{1}{2 \kappa^2} e \, S^2 \,  ,\nonumber \\
&& \mathcal{L}_1 = - \frac{1}{4} e R_{\mu \nu} \, \xi(\Box) R^{\mu \nu} + \frac{1}{8} e R \,  \xi(\Box) R \nonumber \\
&&\hspace{0.9cm} - \frac{1}{4} \epsilon^{\mu \nu \rho} \bar{\psi}_{\mu}   \, \xi(\Box) D^2_{\tau} \mathcal{R}_{\rho \sigma} 
-  \frac{1}{2} e (\partial_{\mu} S) \xi(\Box) (\partial_{\mu} S ) \,  ,
\nonumber \\
&&\mathcal{L}_2 = \frac{1}{32} e R \, \eta(\Box) R 
- \frac{1}{16} e \, \bar{\mathcal{R}}_{\rho \sigma} \gamma^{\rho \sigma} \, \eta(\Box)  \not\!\! D 
\, \gamma^{\tau \lambda} \mathcal{R}_{\tau \lambda} \nonumber \\
&&  \hspace{0.7cm} -  \frac{1}{2} e (\partial_{\mu} S) \eta(\Box) (\partial_{\mu} S ) \,  ,
\ee
where $\mathcal{R}_{\mu \nu} = D_{\mu} \psi_{\nu} - D_{\nu} \psi_{\mu}$. 
To get the propagators around the flat spacetime, we look into the bilinear terms of the total 
Lagrangian $\mathcal{L} = \mathcal{L}_0+\mathcal{L}_1+\mathcal{L}_2$, namely 
\be
&& \hspace{-0.65cm} 
\mathcal{L}_{\rm lin} = \frac{1}{4} h^{\mu \nu} \Big[ \left (\kappa^{-2} - \xi(\Box) \, \Box \right)
P^{(2)}_{\mu \nu, \rho \sigma}
\nonumber \\ 
&&- \left( \kappa^{-2} -(\xi(\Box) +\eta(\Box) ) \, \Box \right) P^{(0,s)}_{\mu \nu, \rho \sigma} \Big] \Box h^{\rho \sigma} \nonumber \\
&& + \frac{1}{2} \bar{\psi}^{\mu} \Big[ \left(  \kappa^{-2} - \xi(\Box)  \, \Box \right)P^{(3/2)}_{\mu \nu} \nonumber \\
&& - \left( \kappa^{-2} -   (\xi(\Box) +\eta(\Box) ) \, \Box \right) (P^{(1/2)}_{11})_{\mu \nu} \Big] \not\! \partial \psi^{\nu} \nonumber \\
&& - \frac{1}{2} S \left( \kappa^{-2} -(\xi(\Box) +\eta(\Box) ) \, \Box \right) S \, , 
\label{Llin3d}
\ee
where the $3D$-projectors are the analog of the corresponding ones in $4D$ \cite{3dSugra}.
Inverting the spin-blocks in (\ref{Llin3d}), we get the propagators for the 
$h_{\mu \nu}$, $\psi^{\mu}$ and $S$ 
fields,
\be
&& \hspace{-0.4cm} 
\langle h_{\mu\nu} h_{\rho \sigma} \rangle \! \propto  \! \frac{1}{\Box} \! \left[
 \frac{P^{(2)}_{\mu \nu, \rho \sigma}}{ 1- \kappa^2 \Box \, \xi(\Box)  }
- \frac{P^{(0,s)}_{\mu \nu, \rho \sigma}}{ 1- \kappa^2 \Box \, (\xi(\Box)+ \eta(\Box))  } \right] \! ,
\nonumber \\
&& \hspace{-0.4cm} 
  \langle \psi_{\mu} \psi_{\nu} \rangle \! \propto \! \frac{1}{\not\!\partial} \left[ 
  \frac{P^{(3/2)}_{\mu \nu}}{ 1-  \kappa^2 \Box \, \xi(\Box)  }
 - \frac{P^{(1/2)}_{\mu \nu}}{ 1-  \kappa^2 \Box \, (\xi(\Box) + \eta(\Box) ) } \right]  \! , 
 \nonumber \\
 && 
 \hspace{-0.4cm} 
 \langle S  S \rangle \! \propto \! \frac{ 1 }{\kappa^2 (\xi(\Box) + \eta(\Box) ) -1} .
 \label{prop3d}
\ee
In analogy with (\ref{alphabeta}), we again introduce two form factors $h_0(z)$ and $h_2(z)$
in the following way,
\be
\xi(\Box) = \alpha_0 + h_0(\Box) \,\, , \,\,\,\, 
\eta(\Box) =  \beta_2 + h_2(\Box).
\ee
To avoid extra poles in the propagators,  we consider the following entire functions,
\be
&& h_0(\Box) :=  - 2 \frac{e^{H(\Box)} - 1 }{\kappa^{-2} \Box} - \tilde{\beta}_0 \,  , 
\nonumber \\
 && h_2(\Box) :=  - \frac{e^{H(\Box)} - 1 }{\kappa^{-2} \Box} - \tilde{\alpha}_2 \,  .
\ee
We assume the quantum theory is renormalized at some scale $\mu_0$ and 
we set  
\be
 {\alpha}_0(\mu_0) = \tilde{\alpha}_0 \,,  \,\,\,\, 
 {\beta}_2(\mu_0) = \tilde{\beta}_2 \, . 
\label{betaalpha3d}
\ee
The propagators in (\ref{prop3d}) simplify to 
\be
&& \hspace{-0.4cm} 
\langle h_{\mu\nu} h_{\rho \sigma} \rangle \! \propto  \! \frac{e^{-H(\Box)}}{\Box} \! \left(
 P^{(2)}_{\mu \nu, \rho \sigma}
- P^{(0,s)}_{\mu \nu, \rho \sigma} \right) \, ,
\nonumber \\
&& \hspace{-0.4cm} 
  \langle \psi_{\mu} \psi_{\nu} \rangle \! \propto \! \frac{e^{-H(\Box)} }{\not\!\partial} \left( 
  P^{(3/2)}_{\mu \nu}  
 - P^{(1/2)}_{\mu \nu} \right)  \, , 
 \nonumber \\
 && 
 \hspace{-0.4cm} 
 \langle S  S \rangle \! \propto \! e^{-H(\Box)} \, .
\ee
The theory respects global supersymmetry,
since there are no other poles in the propagators besides those obtained in the $\mathcal{L}_0$ theory. 

For the $N=2$ supergravity, we proceed in the same way. The multiplet now consists of
$8+8$ degrees of freedom: the graviton, 
a Dirac spinor, a real vector and a complex scalar $(e^m_{\mu}, \psi_\mu, \psi^*_\mu, A_{\mu}, B, B^*)$. 
 The non-zero two point functions are
\be \langle h_{\mu \nu} h_{\rho \sigma} \rangle, \,\,\, \langle \psi_{\mu} \bar{\psi}^*_{\nu} \rangle, \,\,\, 
\langle B B^* \rangle,
\ee
which are identical to those in (\ref{prop3d}), plus the vector field two point function.
The bilinear Lagrangian in the vector field reads 
\be
&& \hspace{-0.8cm}\mathcal{L}(A)_{\rm lin} = \frac{1}{2} A^{\mu} 
\Big[ \left(\kappa^{-2} - \xi(\Box) \Box \right) P_{\mu\nu}^{(T)}  \\
&& \hspace{1.4cm} + \left( \kappa^{-2} - (\xi(\Box) + \eta(\Box) )  \right) P_{\mu \nu}^{(L)} \Big] A^{\nu} \, ,
\nonumber 
\ee
which leads us to the following propagator,
\be
\langle A_{\mu} A_{\nu} \rangle \propto  \frac{P^{(T)}_{\mu \nu}}{1 - \kappa^{2} \xi(\Box) \Box } + 
\frac{P^{(L)}_{\mu \nu}}{1 - \kappa^{2} (\xi(\Box) + \eta(\Box) \Box } \,  ,
\label{A2}
\ee
where $P^{(T)}$ and $P^{(L)}$ are the usual transversal and longitudinal vector field projectors.
The two point function (\ref{A2}), for the same choice of $h_2(z)$ and $h_0(z)$ and the same 
identification (\ref{betaalpha3d}), 
reads 
\be
\langle A_{\mu} A_{\nu} \rangle \propto e^{ - H(\Box)} \left( P^{(T)}_{\mu \nu} + P^{(L)}_{\mu \nu} \right)  ,
\ee
which starts to propagate in the ultraviolet regime like the other auxiliary fields.

In this section we have provided further evidence ($N=1$ and the $N=2$ three dimensional
supergravity)
that non-locality nails the spectrum of the (curvature)$^2$ theory to be the same of linear theory 
while maintaining unitarity.


\begin{acknowledgments}
\noindent Research at Perimeter Institute is supported by the Government of Canada through Industry Canada and by the Province of Ontario through the Ministry of Research \& Innovation.
\end{acknowledgments}

\end{document}